\documentclass[final,5p,times,twocolumn,preprint]{elsarticle}
\biboptions{sort&compress}
\usepackage{amsmath,amssymb,amsfonts,latexsym}
\usepackage[hypertexnames=false, colorlinks=true, citecolor=blue, urlcolor=blue, linkcolor=black]{hyperref}
\usepackage{xcolor}
\usepackage{cleveref}
\usepackage{soul}
\crefname{appendix}{}{}
\allowdisplaybreaks
\newcommand{\FF}{F_\pi^Q}
\newcommand{\sth}{{s_0}}
\begin{document}
\begin{frontmatter}
\title{Pion transverse charge density from $e^+e^-$ annihilation data \\
and logarithmic dispersion relations}
\author[a,b]{Enrique Ruiz Arriola}
\author[a]{Pablo Sanchez-Puertas}
\author[c]{Christian Weiss}
\address[a]{Departamento de F\'isica At\'omica, Molecular y Nuclear,
Universidad de Granada, E-18071, Granada, Spain}
\address[b]{Instituto Carlos I de F\'isica Te\'orica y Computacional,
Universidad de Granada, E-18071, Granada, Spain}
\address[c]{Theory Center, Jefferson Lab, Newport News, VA 23606, USA}
\begin{abstract}
The transverse charge density of the pion is extracted from a dispersive analysis of the 
$e^+e^- \rightarrow \pi^+\pi^-$ exclusive annihilation data. A logarithmic dispersion relation 
is used to compute the unknown phase of the timelike pion form factor from the modulus obtained
from the annihilation cross section. The method is model-independent and permits quantitative
uncertainty estimates. The density is obtained with few-percent accuracy down to $b \sim 0.1$ fm;
at smaller distances it depends qualitatively on the assumed high-energy behavior of the timelike
form factor. Implications for pion structure and the relevance of pQCD asymptotics are discussed.
\end{abstract}
\end{frontmatter}

\section{Introduction}
\label{sec:introduction}
The internal structure of the pion plays a central role in quantum chromodynamics
(QCD) and its application to hadronic systems.
The pion is the Goldstone boson associated with the spontaneous breaking of chiral symmetry,
and its structure is governed by the non-perturbative interactions causing this effect.
The charged pion form factor (FF) and the neutral pion transition FF at momentum
transfers $Q^2 \gg 1\, \textrm{GeV}^2$ are the simplest hadron FFs that can be analyzed
using QCD factorization \cite{Radyushkin:1977gp,Farrar:1979aw,Efremov:1978rn,%
Lepage:1979zb,Efremov:1979qk,Lepage:1980fj}, and their study is an important
testing ground for perturbative QCD (pQCD) interactions \cite{Gross:2022hyw}.

The transverse charge density is a powerful tool for quantifying
pion structure in QCD.  It is defined as the two-dimensional Fourier
transform of the pion FF $\FF (q^2)$ at spacelike momentum transfers
$q^2 = -Q^2 < 0$ and describes the distribution of charge in
transverse position at fixed light-front time
\cite{Soper:1976jc,Burkardt:2000za,Burkardt:2002hr,Miller:2010nz}.  It
provides a concept of spatial density appropriate for the relativistic
system and corresponds to a genuine quantum-mechanical expectation
value of the electromagnetic current in the localized pion state.
In the partonic picture, the transverse density
represents a projection of the generalized parton distributions (GPDs)
in the pion and provides information on the distribution of quarks/antiquarks
in the system \cite{Diehl:2003ny,Belitsky:2005qn,Boffi:2007yc}.
In the hadronic picture, the transverse density
is analyzed using methods based on analyticity and unitarity
and expresses the dynamics of hadronic exchange mechanisms. As such the transverse density
allows one to connect partonic structure with hadronic dynamics and
offers unique insight into hadron structure.

The pion FF at timelike momentum transfers $q^2 = s$ is measured in
exclusive annihilation experiments $e^+e^- \rightarrow
\pi^+\pi^-$ \cite{Druzhinin:2011qd}. Through the dispersion relations for the complex function
$\FF(q^2)$, these data can be used to extract the spacelike FF and
the transverse density \cite{Miller:2010tz}. The standard dispersion relation requires
knowledge of the imaginary part $\textrm{Im} \, \FF(s)$ at $s > 4
M_\pi^2$, while the experiments only provide the modulus
$|\FF(s)|^2$. This necessitates the use of model-dependent methods
(e.g.\ fits by a superposition of resonances \cite{Bruch:2004py}) to
reconstruct the phase of the FF. An alternative approach is the use of
logarithmic (or phase--modulus) dispersion relations, which
reconstruct the phase of the FF from the $s$-dependence of the
measured $|\FF(s)|^2$, provided the FF has no complex zeros
\cite{Truong:1969gr,Geshkenbein:1998gu}.  The method is
unique to the pion FF, for which $|\FF(s)|^2$ can be measured starting
from the threshold at $s = 4 M_\pi^2$, and becomes practical thanks to
the high precision of the annihilation data.  It offers several
advantages over the conventional dispersive analysis: it avoids
model dependence, works directly with experimental data, and permits
quantitative uncertainty estimates. An extraction of the phase of the
pion FF using logarithmic dispersion relations was performed recently in
Ref.~\cite{RuizArriola:2024gwb} (see also \cite{Leplumey:2025kvv}).

In this work we analyze the pion transverse charge density using
logarithmic dispersion relations and the results of
Ref.~\cite{RuizArriola:2024gwb}.  We extract the empirical density
over a wide range of distances and compare its
behavior with theoretical expectations based on pQCD, vector
resonances and unitarity, and chiral dynamics. Our approach offers
several new aspects: (i) It represents a unique case of ``transverse
imaging'' based directly on experimental data.  (ii) It permits simple
and efficient statistical uncertainty quantification of the transverse density,
avoiding issues with Fourier transform of spacelike FF data.  (iii) It
constrains the density at distances $\gtrsim 0.1$~fm with smaller fit
uncertainties than other methods. (iv) It provides
an estimate of the dominant theoretical uncertainty based on QCD arguments.

\section{Transverse density and dispersive representation}
The transverse charge density is defined as the 2-dimensional Fourier transform 
(radial Fourier--Bessel transform) of the spacelike pion FF,
\begin{align} 
\rho_\pi (b) \;\; = \;\; 
\int\limits_0^\infty \frac{dQ}{2\pi} \, 
Q \, J_0 (Qb) \; \FF(-Q^2).
\label{rho_def}
\end{align}
It describes the distribution of charge with respect to the distance $b$ from the 
transverse center of momentum, with $\int d^2 b \, \rho_\pi (b) = 1$; for its
interpretation and connection with GPDs, see 
Refs.~\cite{Burkardt:2000za,Burkardt:2002hr,Miller:2010nz}.
Using the analytic properties one can express the density in terms of the FF
in the timelike region. The pion FF satisfies an unsubtracted dispersion relation,
\begin{align} 
\FF(-Q^2) \;\; = \;\; \int\limits_\sth^\infty ds \, \frac{\textrm{Im}\, \FF(s)}{\pi (s + Q^2 - i0)},
\label{disp_ff}
\end{align} 
where $\textrm{Im}\, \FF(s)$ is the imaginary part on the cut at $s > \sth \equiv 4 M_\pi^2$.
Substituting this representation into Eq.~(\ref{rho_def}) and performing the Fourier integral one obtains
\cite{Strikman:2010pu}
\begin{align} 
\rho_\pi (b) \;\; = \;\; \int\limits_\sth^\infty \frac{ds}{2\pi^2} 
\; K_0(\sqrt{s} b) \; \textrm{Im}\, \FF(s) ,
\label{disp_dens}
\end{align}
where $K_0$ is the modified Bessel function.
This dispersive representation of the density has several
useful properties: (i)~The kernel decays exponentially at large arguments,
$K_0 (x) \sim \sqrt{\pi / 2} \exp (-x)/\sqrt{x}$ for $x \gg 1$, so
that the dispersion integral converges exponentially at large energies,
to be compared with the oscillating behavior of $J_0$ in Eq.~(\ref{rho_def}).
(ii)~The range of $\sqrt{s}$ covered in the integral is determined
by the inverse distance, $1/b$, and the density can be regarded as an ``exponential filter''
sampling the timelike FF.\footnote{This property is similar to the so-called
Laplace sum rules for Euclidean correlation functions \cite{Narison:1981ts,Narison:2023ntg}.}
(iii)~The kernel is
positive, $K_0 > 0$, simplifying the uncertainty propagation from the
pion FF into the density.
(iv)~The dispersive representation embodies the
analytic properties of the FF and produces densities with the correct
asymptotic behavior at $b \rightarrow \infty$, enabling mathematically
stable computation of densities at large distances
\cite{Alarcon:2022adi}.
\section{Basic properties and asymptotic behavior}
\label{sec:properties}
Several properties of the transverse density can be derived directly
from the dispersive representation Eq.~(\ref{disp_dens})
and will be of relevance in what follows.

\textit{Large distances and threshold behavior.}
At distances $b \gg M_\pi^{-1}$
the dispersion integral is dominated by energies near the threshold,
$s = \sth + \textrm{few} \, M_\pi^2$.
Below the $4\pi$ threshold at $s = 16   M_\pi^2$, the
phase of the pion FF is determined by elastic unitarity (Watson's theorem),
\begin{align}
\FF(s)= |\FF(s)| \, e^{i \delta_1^1(s)}, \hspace{2em}
\textrm{Im} \FF(s) = |\FF(s)| \sin \delta_1^1(s),
\label{spectral_unitarity}
\end{align}
where $\delta_1^1$ is the phase shift of $\pi\pi$ scattering in the $I
= J = 1$ isospin and angular momentum channel.
In terms of the partial-wave amplitude (in the convention of Ref.~\cite{Gasser:1983yg})
\begin{align}
f_1^1 (s) \equiv e^{i \delta_1^1 (s)} \sin
\delta_1^1 (s) \sqrt{s}/(2 k_\pi),
\label{f_from_delta}
\end{align}
where $k_\pi \equiv \sqrt{s/4 - M_\pi^2}$ is the $\pi\pi$ CM momentum, the threshold expansion
is given by
\begin{align}
  {\rm Re} f_1^1 (s) = \frac{\sqrt{s} }{ 4 k_\pi}
  \sin ( 2\delta_1^1 (s))= 2 M_\pi k_\pi^2 \left( a_{1}^1 + b_{1}^1 k_\pi^2 + \dots \right) \, , 
\label{f_threshold}
\end{align}
where $a_1^1$ and $b_1^1$ are the slope parameters. For the
imaginary part this implies
\begin{align}
{\rm Im} \FF (s) =  |\FF (4M_\pi^2)| \, a_1^1 \, k_\pi^3 + \mathcal{O}(k_\pi^5). 
\label{spectral_threshold}
\end{align}
The asymptotic behavior of the density is then obtained from Eq.~(\ref{disp_dens}) as
\begin{align}
\rho_\pi (b) \sim \frac{3 M_\pi^2 \, |\FF(4M_\pi^2)| \, a_{1}^1 }{2 \pi b^3} \,
e^{-2 M_\pi b} \hspace{2em} (b \gg M_\pi^{-1}).
\label{large_b-threshold}
\end{align}
Chiral perturbation theory provides approximate predictions for the threshold parameters.
At LO accuracy \cite{Gasser:1983yg}
\begin{align}
a_1^1 =2/(96\pi F_{\pi}^2 M_{\pi}) + \dots, \hspace{2em}
|\FF (4M_\pi^2)| = 1 + \dots ,
\label{chpt}
\end{align}
where $F_\pi =$ 92 MeV is the pion decay constant,
and the asymptotic behavior Eq.~(\ref{large_b-threshold}) becomes
\begin{align}
\rho_\pi (b) \sim
\frac{3 M_\pi}{96 \pi^2 F_\pi^2 b^3}  \, e^{-2 M_\pi b} \hspace{2em} (b \gg M_\pi^{-1}) .
\label{large_b}
\end{align}
At what values of $b$ the asymptotic behavior Eq.~(\ref{large_b-threshold}) becomes relevant,
and how accurately the absolute prediction Eq.~(\ref{large_b}) describes the asymptotic density,
are questions that can be answered by the empirical results.

\textit{Intermediate distances and vector meson dominance.}  The inelasticity in $\pi\pi$ scattering
remains small up to the $K\bar K$ threshold \cite{Colangelo:2001df,Garcia-Martin:2011iqs},
and the elastic unitarity approximation Eq.~(\ref{spectral_unitarity}) for the timelike FF
can be used up to $s_1 = 4 M_K^2 \approx 1\, \textrm{GeV}^2$ \cite{Druzhinin:2011qd}.
For the density this implies
\begin{align}
\rho_\pi (b) \approx \int_\sth^{s_1} \frac{ds}{2\pi^2} K_0(\sqrt{s} b) |\FF(s)| \sin \delta_{1}^1(s)
+ {\cal O} (e^{- \sqrt{s_1} b}).
\label{density_elastic}
\end{align}
In this  region the phase shift is dominated by the $\rho$ resonance, and the density decays
exponentially with the $\rho$ meson mass. A zero-width resonance, with a coupling chosen
such that $\FF (0) = 1$, corresponds to $\textrm{Im} \FF(s) = \pi M_\rho^2 \delta (s - M_\rho^2)$
in Eq.~(\ref{disp_ff}) and provides an approximation to the density in Eq.~(\ref{disp_dens}) as
\begin{align}
\rho_\pi (b)
= (M_\rho^2/2\pi) \, K_0 (M_\rho b).
\label{rho_zero_width}
\end{align}
This dynamics governs the behavior of density at intermediate distances
$b \sim 1/M_\rho$ and can be seen in the empirical results.

\textit{Small distances and pQCD.}
The asymptotic behavior of the density at $b \rightarrow 0$ can be derived from Eq.~(\ref{disp_dens})
using the pQCD approximation for the timelike pion FF. The asymptotic behavior of the
spacelike pion FF is given by the well-known expression resulting from the $\mathcal{O}(\alpha_s)$
hard scattering process and the asymptotic pion distribution amplitudes
\cite{Radyushkin:1977gp,Farrar:1979aw,Efremov:1978rn,Lepage:1979zb,Efremov:1979qk,Lepage:1980fj},
\begin{align}
\FF (-Q^2) = \frac{16\pi F_\pi^2 \alpha_s (Q^2)}{Q^2} \left[ 1 + \mathcal{O}(\alpha_s) \right] ,
\label{eq:pQCD_spacelike}
\end{align}
where the running coupling $\alpha_s$ is taken at the scale $Q^2$. At LO accuracy the coupling
is given by 
$\alpha_s(Q^2) = 4\pi/ [\beta_0 \log (Q^2 / \Lambda_{\rm QCD}^2)]$,
where $\beta_0 = (11N_c - 2 N_f)/3$ and $\Lambda_{\rm QCD}$ is the QCD scale parameter.
Analytic continuation to timelike momenta is performed by setting $Q^2 = -s - i0$
and continuing from $s < 0$ to $s > 0$, resulting in
\begin{align}
\alpha_s \rightarrow
\frac{4\pi}{\beta_0 [\ln(s/\Lambda^2_{\textrm{QCD}}) -i\pi]} .
\label{alphas_timelike_lo}
\end{align}
The imaginary part of the pion FF in LO is thus given by
\begin{align}
\textrm{Im} \, \FF (s) = -\frac{64 \pi^3 F_\pi^2}
{\beta_0 s \, [ \log^2 (s/\Lambda_{\rm QCD}^2)+\pi^2]} .
\label{eq:pQCD-LO}
\end{align}
The negative sign unambiguously follows from the analytic continuation.
Because at low and intermediate energies $\operatorname{Im}\FF (s) > 0$,
see Eqs.~(\ref{spectral_threshold}) and (\ref{chpt}), Eq.~(\ref{eq:pQCD-LO}) implies that
$\operatorname{Im}\FF (s)$ must change sign as a function of $s$.
This is indeed seen in the empirical timelike FF (see Fig.~\ref{fig:tImF}) \cite{RuizArriola:2024gwb}.

The imaginary part of the pion timelike FF obeys the sum rules
\cite{Donoghue:1996bt,Sanchez-Puertas:2024siv}
\begin{align}
\frac1{\pi }\int_\sth^\infty ds\, \frac{\textrm{Im}\,  \FF(s)}{s} = 1,
\label{sumrule_charge}
\\
\frac1{\pi } \int_\sth^\infty  ds\, \textrm{Im}\, \FF(s) = 0.
\label{sumrule_asymptotic}
\end{align}
The first relation follows from the pion charge, $\FF (0) = 1$.
The second relation expresses the absence of a $1/Q^2$ power term
in the asymptotic expansion of the spacelike FF at large $Q^2$,
as required by pQCD. Note that the large-$s$ behavior of $\textrm{Im} \, \FF(s)$
in pQCD, Eq.~(\ref{eq:pQCD-LO}), is such that the integral
Eq.~(\ref{sumrule_asymptotic}) converges. The sum rules
Eqs.~(\ref{sumrule_charge}) and (\ref{sumrule_asymptotic}) by themselves
imply that $\operatorname{Im}\FF(s)$ must change sign as a function of $s$,
as observed in the explicit expressions above.

The pQCD prediction for the imaginary part, Eq.~(\ref{eq:pQCD-LO}), combined with the sum rule
Eq.~(\ref{sumrule_asymptotic}), implies a certain asymptotic behavior of the density at
small distances. Expanding the kernel in Eq.~(\ref{disp_dens}),
$K_0(x)= -\log (x) -\log(e^{\gamma_E}/2) + {\cal O} [x^2 \log(x)]$ for $x \ll 1$,
we obtain
\begin{align}
\label{eq:AsympRhoB}
\rho_\pi (b) = \int_\sth^\infty \frac{ds}{2\pi^2}
\left[ -\log (\sqrt{s}b) -\log (e^{\gamma_E}/2)  \right] \,
\textrm{Im}\, \FF(s) .
\end{align}
Separating the arguments in the first logarithm, the coefficient of $-\log b -\log(e^{\gamma_E}/2)$ vanishes
due to Eq.~(\ref{sumrule_asymptotic}) (this finding will have important implications later).
The asymptotic behavior of $\rho_\pi (b)$ results from the $-\log\sqrt{s}$ term and is computed
by restricting the integral to values $\sqrt{s} b \ll 1$, giving
\begin{align}
\rho_\pi (b) \sim
\frac{16 \pi F_\pi^2 }{\beta_0} \left[ \log \log (1/b \Lambda_{\rm QCD} ) + c \right]
\, + \, {\cal O} [1/\log (b)],
\label{eq:rho-short}
  \end{align}
where $c$ is a constant that governs the numerical approach to the double logarithmic
asymptotic behavior; the choice of $\Lambda_{\rm QCD}$ as mass scale in the double
logarithm is arbitrary and determines the value of the constant.
The asymptotic behavior Eq.~(\ref{eq:rho-short}) agrees with the one derived from
the spacelike FF in Ref.~\cite{Miller:2009qu}. It is derived from
pQCD and not expected to be applicable at practically achievable energies,
as seen in the empirical results; see Refs.~\cite{RuizArriola:2008sq,Ananthanarayan:2012tn}
for further discussion. In particular, Eq.~(\ref{eq:rho-short}) relies on the asymptotic
sum rule Eq.~(\ref{sumrule_asymptotic}); any nonzero value of the integral would produce
a $\log (1/b)$ asymptotic behavior, which would overrule that of Eq.~(\ref{eq:rho-short}).

\section{Imaginary part from logarithmic dispersion relation}  
To evaluate the density in the dispersive representation Eq.(\ref{disp_dens}) one needs the imaginary
part of the pion FF over a broad range of energies.
Exclusive annihilation experiments $e^+e^- \rightarrow \pi^+\pi^-$
measure the squared modulus $|\FF(s)|^2$ starting from the threshold $s = \sth$. The phase of the FF
can be reconstructed from the $s$-dependence of the modulus using logarithmic dispersion
relations. To derive them one sets $\FF(s) = |\FF(s)| e^{i\delta (s)}$, assumes that $\FF(s)$
has no zeroes in the complex plane,
and writes a dispersion relation for the complex function $\log \FF (s)/(\sth -s)^{n + 1/2}$
with $n = 0, 1, ...$, which has the same cut as the original $\FF(s)$.
The dispersion relation for $n = 1$ is obtained as \cite{RuizArriola:2024gwb}
\begin{align} 
\delta(s) &= \frac{-s (s-\sth )^{3/2}}{2\pi} \; \textrm{P}
\int_\sth^{\infty}ds' \frac{\log|\FF(s')/\FF(\sth )|^2}{s'(s'-s)(s'-\sth )^{3/2}}
\nonumber\\
&-(s/\sth -1)^{3/2}\log \FF(\sth ) ;
\label{eq:DR2}
\end{align}
it incorporates the condition $\FF(0)=1$ and implements the correct $P$-wave threshold behavior.
If the variation of the logarithm is limited,
$\log |\FF(s')/ \FF (\sth )| \sim \textrm{const}$, the integrand in Eq.~(\ref{eq:DR2}) behaves 
as $\sim (s')^{-7/2}$ at large $s'$, providing rapid convergence. Eq.~(\ref{eq:DR2}) is therefore
well suited for numerical evaluation.

An extraction of the phase of the pion FF using Eq.~(\ref{eq:DR2}) was performed
in Ref.~\cite{RuizArriola:2024gwb}. The analysis used the BaBar data \cite{BaBar:2009wpw} and
obtained the phase with controlled uncertainties up to $s_{\rm max} = (2.5~\textrm{GeV})^2$;
see Ref.~\cite{RuizArriola:2024gwb} for discussion of the data selection.
Various methods were employed to interpolate the data at $s > (0.6~\textrm{GeV})^2$
(Gounaris-Sakurai parametrization \cite{Gounaris:1968mw}, linear interpolation)
and produced consistent results. Uncertainties were estimated using a MC procedure,
fully accounting for correlations without bias. The method enables propagation of the uncertainties into
functions derived from the phase, such as the transverse charge density.

%
%
\begin{figure}[t]
\includegraphics[width=0.48\textwidth]{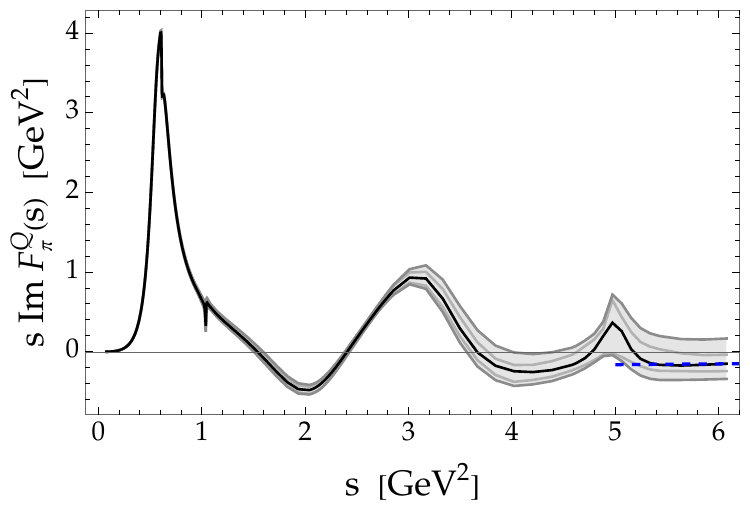}
\caption{Imaginary part of the pion FF, $s \operatorname{Im} \FF(s)$,
extracted from the logarithmic dispersion relation analysis of Ref.~\cite{RuizArriola:2024gwb}.
\textit{Black line:} Central value.
\textit{Inner gray band:} Statistical uncertainty.
\textit{Outer gray band:} Total uncertainty.
\textit{Dashed blue line:} pQCD prediction Eq.~(\ref{eq:pQCD}) (details in text).}
\label{fig:tImF}
\end{figure}
Figure~\ref{fig:tImF} shows the imaginary part of the FF obtained from the analysis of
Ref.~\cite{RuizArriola:2024gwb}.
The inner band represents the statistical uncertainty estimated with a MC procedure; 
the outer band includes also the interpolation uncertainty. One observes:
(i)~At $s \lesssim 1$ GeV$^2$ the imaginary part is dominated by the $\rho$ resonance, as predicted
by elastic unitarity, Eq.~(\ref{spectral_unitarity}).
(ii)~At higher $s$ the imaginary part exhibits a pattern of resonances with alternating sign,
with the $\rho'$ and $\rho''$ clearly established and two further peaks suggested.
The same pattern was obtained in a resonance-based fit to the timelike FF data \cite{Bruch:2004py}
and a dispersive analysis of the spacelike data \cite{RuizArriola:2008sq} and is
well-established in the dispersive approach.
(iii)~The behavior at large $s$ appears to be qualitatively consistent with an approach to a negative
value as predicted by pQCD; see the LO expression in Eq.~(\ref{eq:pQCD-LO}).
The dashed blue line shows $\textrm{Im}\, \FF(s)$
corresponding to the NNLO prediction for the timelike FF
\cite{Chen:2023byr,Ji:2024iak}\footnote{We use the updated arXiv version of Ref.~\cite{Chen:2023byr},
which matches the recent calculation in Ref.~\cite{Ji:2024iak} when accounting for scheme- and
scale dependencies.  We assume an asymptotic distribution amplitude and factorization scale $\mu=Q$.}
\begin{align}
\FF(s) = -\frac{16\pi F_{\pi}^2 \alpha_s}{s}
\left(1 + 6.58\frac{\alpha_s}{\pi} \left[ 1 + 8.69\frac{\alpha_s}{\pi} \right] \right),
\label{eq:pQCD}
\end{align}
where $\alpha_s$ is the timelike coupling evaluated with the LO expression
Eq.~(\ref{alphas_timelike_lo}) with $\Lambda_{\rm QCD} = 250$ MeV; the effect of
higher-order corrections to $\alpha_s$ is negligible compared to the corrections
to the FF in Eq.~(\ref{eq:pQCD}).
The approach of the empirical FF to the pQCD prediction should be understood on average,
as suggested by the Phragmen-Lindel{\"o}f theorem for the asymptotic behavior of
complex functions with branch cut singularities \cite{Truong:1969gr} and the
phenomenology of quark-hadron duality. For further discussion of the asymptotic
behavior of $\textrm{Im}\, \FF(s)$ in pQCD, see Refs.~\cite{Gousset:1994yh,Chen:2018tch}.

For the following it will be important to quantify the convergence of the sum rules
Eqs.~(\ref{sumrule_charge}) and (\ref{sumrule_asymptotic}). Integrating the empirical
$\textrm{Im}\, \FF(s)$ over $s$ up to $s_{\rm max}$ we obtain
\begin{align}
\frac{1}{\pi} \int_{\sth}^{s_{\rm max}} ds
\frac{\operatorname{Im} \FF (s)}{s} \Big|_{\rm Data}
&= 1.01(1)_{\textrm{st}}(^{+2}_{-1})_{\textrm{syst}},
\label{sumrule_charge_partial} \\
\frac{1}{\pi} \int_{\sth}^{s_{\rm max}} ds \operatorname{Im} \FF (s) \Big|_{\rm Data}
&= 0.63(2)_{\textrm{st}}(^{+7}_{-4})_{\textrm{syst}}~\textrm{GeV}^2.
\label{sumrule_asymptotic_partial}
\end{align}
The partial integral Eq.~(\ref{sumrule_charge_partial}) satisfies the charge sum rule Eq.~(\ref{sumrule_charge})
within uncertainties. The partial integral Eq.~(\ref{sumrule_asymptotic_partial}) has a significant non-zero value
and is far from satisfying the asymptotic sum rule Eq.~(\ref{sumrule_asymptotic}); note that the natural mass scale
of the dimensionful integral is $M_\rho^2 \approx$ 0.6 GeV$^2$. The asymptotic sum rule thus requires substantial
contributions from energies $s > s_{\rm max}$. An interesting question is whether the pQCD result
could account for this missing contribution. Integrating $\textrm{Im}\, \FF(s)$ of
Eq.~(\ref{eq:pQCD}) from $s_{\rm max}$ to infinity we obtain (at LO, NLO, NNLO accuracy)
\begin{align}
\frac{1}{\pi} \int_{s_{\rm max}}^\infty ds
\frac{\operatorname{Im} \FF (s)}{s}\Big|_{\rm pQCD}
&= - \underbrace{0.0025}_{\rm LO}- \underbrace{0.0011}_{\rm NLO}- \underbrace{0.0006}_{\rm NNLO} \, , 
\label{sumrule_charge_pqcd} \\
\frac{1}{\pi} \int_{s_{\rm max}}^\infty ds \operatorname{Im}\FF (s) \Big|_{\rm pQCD}
&= - \underbrace{0.114}_{\rm LO}- \underbrace{0.030}_{\rm NLO}- \underbrace{0.013}_{\rm NNLO} \, \textrm{GeV}^2.
\label{sumrule_asymptotic_pqcd}
\end{align}
Adding these contributions to Eqs.~(\ref{sumrule_charge_partial}) and (\ref{sumrule_asymptotic_partial})
leaves intact the charge sum rule Eq.(\ref{sumrule_charge}) but is by far not sufficient to fulfill
the asymptotic sum rule Eq.(\ref{sumrule_asymptotic}). Even lowering the lower limit of the integral
in Eq.~(\ref{sumrule_asymptotic_pqcd}) to $s_{\rm max} = 1$ GeV$^2$, the pQCD result
can not account for the missing strength (see also Ref.~\cite{Sanchez-Puertas:2024siv}).
The origin of the missing strength in
the asymptotic sum rule, and indeed the status of the sum rule itself, therefore remain unclear.
These circumstances need to be taken into account in the uncertainties
of the transverse density.
\section{Transverse density and uncertainties}
%
%
\begin{figure}
\includegraphics[width=\linewidth]{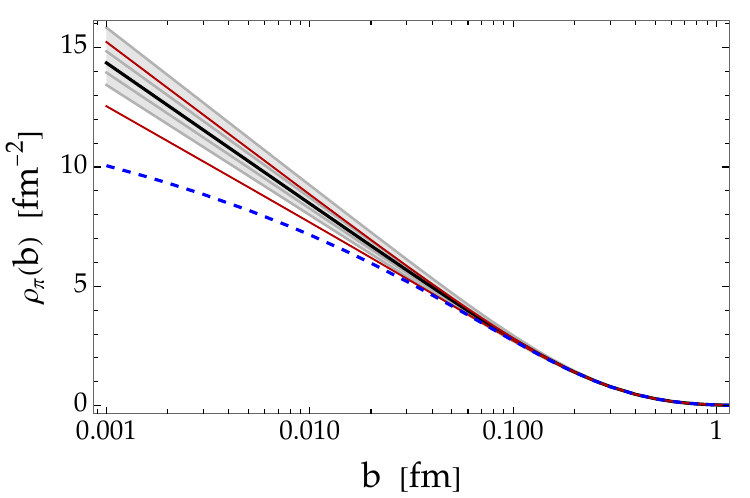}
\caption{Pion transverse charge density $\rho_\pi (b)$ computed with
the imaginary part of pion FF extracted from the logarithmic dispersion relation
(see Fig.~\ref{fig:tImF}) and various high-energy completions.
\textit{Solid black line:} Density from dispersion integral with
finite cutoff $s_{\rm max} = (2.5~\textrm{GeV})^2$.
\textit{Inner and outer gray bands:} Statistical and total uncertainty
from data below $s_{\rm max}$.
\textit{Solid red lines:} Density with high-energy completion Eq.~(\ref{extension})
with $\epsilon = 1$ (asymptotic sum rule disregarded).
\textit{Dashed blue line:} Same with $\epsilon = 0.11$ (asymptotic sum rule satisfied).
\label{fig:density}}
\end{figure}
We now evaluate the transverse density in the dispersive representation Eq.~(\ref{disp_dens})
and estimate its uncertainties. They arise from the uncertainty of the imaginary part in
the region $s < s_{\rm max}$, and the possible contributions from $s > s_{\rm max}$.

\textit{Small distances.} Figure~\ref{fig:density} shows the extracted density at short distances.
The black line gives the density obtained by numerical
evaluation of the dispersion integral Eq.~(\ref{disp_dens}) with the upper limit
$s_{\rm max}$. The inner gray band shows the uncertainty of the density due to
the statistical uncertainty of the imaginary part (see Fig.~\ref{fig:tImF}).
The outer gray band shows the uncertainty of the density due to the total
(statistical and systematic) uncertainty of the imaginary part. The statistical uncertainties
were computed by averaging over the MC ensemble, where each instance satisfies
analyticity and the logarithmic dispersion relation Eq.~(\ref{eq:DR2});
the method takes into account correlations in the data and propagates them into
the density (a similar procedure was used for nucleon densities in Ref.~\cite{Alarcon:2022adi}).
One observes:
(i)~The density exhibits an approximate $\log (1/b)$ growth
at small $b$. This is a direct consequence of the nonzero value of
the integral Eq.~(\ref{sumrule_asymptotic_partial}), which together with Eq.~(\ref{eq:AsympRhoB})
implies a short-distance behavior as
$\rho_{\pi}(b) ~ \sim [\textrm{integral Eq.~(\ref{sumrule_asymptotic_partial})}] \times \log(1/b)/(2\pi)$.
(ii)~The relative uncertainties of the density grow with decreasing $b$,
as the dispersion integral becomes more sensitive to higher $s$, where data
have large uncertainties.

To estimate the uncertainty coming from the region $s > s_{\rm max}$,
we need to make assumptions about the high-energy behavior of the
imaginary part.  Naively, one expects that this uncertainty becomes
relevant at distances below $b \sim 1/\sqrt{s_{\rm max}} =$ 0.08
fm. In view of the principal questions regarding the high-energy
behavior (onset of pQCD behavior, status of asymptotic sum rule), we
content ourselves with simple models realizing different
scenarios. These models are used only to estimate the incompleteness
error by enforcing the sum rules,
not to predict the actual density (see e.g. Refs.~\cite{RuizArriola:2024uub,Broniowski:2024mpw}).

A minimal model of $\operatorname{Im}\FF (s)$ at $s > s_{\rm max}$ is to extend the extracted values
at $s_{\rm max}$ with a (fractional) power behavior,
\begin{align}
\operatorname{Im}\FF (s) &=
\operatorname{Im}\FF (s_{\rm max}) \left(\frac{s_{\rm max}}{s}\right)^{1+\epsilon},
\quad \epsilon > 0 .
\label{extension}
\end{align}
As the first scenario we take $\epsilon=1$, which would be obtained
from a vector dominance model of the FF with constant width,
disregarding the asymptotic sum rule.  Taking the extreme values of
the extracted imaginary part at the maximum energy, $\operatorname{Im}
\FF(s_{\rm max}) = \{ -0.056, 0.027\}$, and extending them with
Eq.~(\ref{extension}), we compute the densities combining the
contributions from $s < s_{\rm max}$ and $s > s_{\rm max}$. The solid
red lines in Fig.~\ref{fig:density} show the resulting range. One
observes: (i) The high-energy extension has negligible effect at $b >
0.01$ fm. (ii) The densities obtained with this high-energy extension
continue the $\log (1/b)$ growth at small $b$ observed in the $s <
s_{\rm max}$ partial result, which follows from the violation
of the sum rule Eq.~(\ref{sumrule_asymptotic})
and cannot prevail down to arbitrarily small distances.

An alternative scenario is to impose the asymptotic sum rule
Eq.~(\ref{sumrule_asymptotic}).  For this we take
Eq.~(\ref{extension}) with the central value $\operatorname{Im}\FF
(s_{\rm max}) = -0.03497$ and choose $\epsilon$ such that the sum rule
is satisfied, which gives $\epsilon =$ 0.11. This describes an
imaginary part with a strong power-like high-energy tail of negative
sign.\footnote{A similar power-like tail is obtained in the dual
  resonance model of Ref.~\cite{Dominguez:2001zu}. Unfortunately this
  model cannot be matched directly with the empirical result, because
  the alternating sign of resonances displayed in Fig.~\ref{fig:tImF}
  is not reproduced in the model \cite{RuizArriola:2008sq}.}  The
dashed blue line in Fig.~\ref{fig:density} shows the density obtained with
this high-energy extension. One observes: (i) The small-$b$ behavior
of the density is now softer, and the $\log (1/b)$ growth is absent, in
accordance with the findings of Sec.~\ref{sec:properties}.
With the sum rule now satisfied, Eq.~(\ref{eq:AsympRhoB}) implies
a behavior as $\rho_{\pi}(b) \sim c +\mathcal{O}[b^{2\epsilon}\log(1/b)]$
for $\epsilon>0$.  (ii) The high-energy behavior in this scenario
affects the density at distances $b <$ 0.1 fm, in agreement with the
naive estimate.

\begin{figure}
\includegraphics[width=\linewidth]{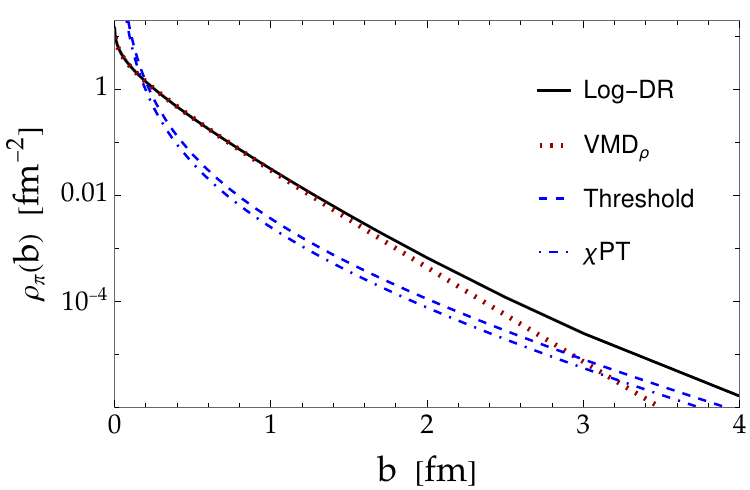}
\caption{Pion transverse charge density at large distances.
\textit{Solid line:} Result of logarithmic dispersion relation (uncertainties not visible on the
scale shown here).
\textit{Dotted red line:} Density from zero-width $\rho$ meson pole Eq.~(\ref{rho_zero_width}).
\textit{Dashed blue line:} Asymptotic behavior from threshold expansion, Eq.~(\ref{large_b-threshold}).
\textit{Dashed-dotted blue line:} Asymptotic behavior from LO chiral perturbation theory, Eq.~(\ref{large_b}).
\label{fig:density_large}
}
\end{figure}

The power-like extension Eq.~(\ref{extension}) does not satisfy the QCD asymptotic behavior.
One could amend the parametrization to include a pQCD tail above some energy $s_{\rm pQCD} > s_{\rm max}$
and adjust the parameters such that both sum rules are satisfied.
The pQCD tail obtained in this way is found to have negligible effect
on the density at the distances $b > 10^{-3}$ fm shown in Fig.~\ref{fig:density_large}.
In particular, the $\log \log (1/b \Lambda_{\rm QCD})$ term in Eq.~(\ref{eq:rho-short}) becomes numerically
comparable to the constant term $c$ only at extremely small distances $b \lesssim 10^{-10}$ fm.
The double logarithmic asymptotics thus plays no role
at realistically accessible distances.

In summary, the density is determined by our analysis down to
distances $b \sim 0.1$ fm with controlled uncertainties.  Its behavior
at smaller distances depends on the assumed high-energy
behavior of imaginary part.  If the asymptotic sum rule is
disregarded, the density is accurately determined by the dispersion
relation down to $b \sim 10^{-3}$ fm, with relative uncertainties
reaching $\sim$20\% at the lower end.  The $\log (1/b)$ rise, first
observed in the analysis of spacelike FF data \cite{Miller:2009qu},
continues down to these distances. If the asymptotic sum rule is
imposed, and the required negative contributions are assumed to come
from energies directly above the measured region, the behavior of the
density is qualitatively different. The rise of the density is slowed
down; however no indications of $\log\log (1/b)$ asymptotic behavior
can be expected over the range covered here.

%
%
\textit{Large distances.}
Figure~\ref{fig:density_large} shows the extracted density at large distances
on a logarithmic scale (the uncertainties are not visible on this scale).
At distances $b \gtrsim 1/M_\rho$ the contributions from energies $s > s_{\rm max}$ are negligible,
and the uncertainty of $\rho_\pi(b)$ arises solely from the statistical and systematic errors of the
dispersion integral. One observes:
(i)~The empirical density agrees very well with the result of
the zero-width $\rho$ meson pole, Eq.~(\ref{rho_zero_width}), up to $b \sim 2$ fm.
(ii)~ The asymptotic behavior obtained from the threshold expansion, Eq.~(\ref{large_b-threshold}),
approximates the density only at very large distances $b > 4$ fm, where it is negligibly small. 
(ii)~The LO chiral perturbation theory result Eq.~(\ref{large_b}) provides a reasonable estimate
of the full threshold expansion result Eq.~(\ref{large_b}).
\section{Implications for partonic structure}
The results for the transverse charge density have interesting implications for the
partonic structure of the pion.

\textit{Positivity and attractive $\pi\pi$ interactions.}
In the partonic picture the transverse charge density is given by
\begin{align} 
\rho_\pi (b) = \int dx \left\{ e_u [f_u - f_{\bar u}](x, b) + e_d [f_d - f_{\bar d}](x, b) \right\} ,
\label{rho_partonic}
\end{align}
where $\{e_u, e_d\} = \{\frac{2}{3}, -\frac{1}{3}\}$ are the quark
charges and $f_i (x, b) (i = u, \bar u, d, \bar d)$ are
the quark/antiquark densities depending on the light-front momentum
fraction $x$ and transverse position $b$, obtained as the transverse
Fourier transform of the GPDs \cite{Burkardt:2000za,Burkardt:2002hr}.
The functions are particle number densities (probabilities)
and individually positive, $f_i(x, b) > 0$ \cite{Burkardt:2001ni,Diehl:2002he}.
At distances of the order of the typical hadronic size of the pion,
$b \sim R_{\rm had} \sim 1/M_\rho$, 
the density in Eq.~(\ref{rho_partonic}) is dominated by the $u$
and $\bar d$ valence quark densities and thus positive, $\rho_\pi (b) > 0$.
In the dispersive representation the density at these distances can be computed
through the elastic unitarity formula in terms of the $\pi\pi$ phase shift $\delta_1^1 (s)$,
Eq.~(\ref{density_elastic}). The positivity of the density then requires
that $0 < \delta_1^1 (s) < \pi$, i.e., that the interactions in the $\pi\pi$
system be attractive. This represents a statement about the hadronic picture
that is obtained from matching with the partonic picture.

\textit{$\rho$ meson exchange and partonic diffusion.} The empirical charge density in
the pion is very well described by $\rho$ meson exchange in the hadronic picture
(see Fig.~\ref{fig:density_large}). In the partonic picture the transverse distribution
arises from the transverse motion of the partons, which involves parton splitting
and develops characteristics of a diffusion process \cite{Gribov:1973jg,Capella:1992yb}.
Exploring the partonic dynamics that produces the transverse profile of
$\rho$ meson exchange is an interesting direction for further study, closely connected
with the notion of parton-hadron duality.

\textit{Small-size $q\bar q$ configurations in pion.}  In the parton
picture the pion is described as a quantum-mechanical system in
quark/gluon degrees of freedom, containing configurations with varying
particle number and spatial size. The transverse charge density represents
a probability density of the wave function and thus provides information
on the distribution of configurations in the system.
The theoretical analysis of Ref.~\cite{Miller:2010tz}
shows that a rising charge density at $b \ll R_{\rm had}$ requires
the presence of small-size $q\bar q$ configurations in the pion
(minimal Fock component, non-exceptional momentum fractions,
transverse sizes $r \ll R_{\rm had}$), and that it cannot be explained
by so-called end-point configurations (multiparticle Fock components,
exceptional momentum fractions $\rightarrow 1$, transverse size $r
\sim R_{\rm had}$). The large empirical density at $b \ll 0.1$ fm
(in either scenario) implies a substantial probability of
small-size $q \bar q$ configurations in the pion \cite{Miller:2010tz}.
Expressing the scenarios for pQCD asymptotics and the $b \rightarrow 0$
limit of the density in the language of partonic configurations
presents an interesting problem for further study.
\section{Conclusions and extensions}
The logarithmic dispersion relation for the pion FF allows one to extract
the transverse charge density from the cross section of $e^+e^- \rightarrow \pi^+\pi^-$
exclusive annihilation, a unique case of ``transverse imaging'' based
directly on experimental data. Present data up to $\sqrt{s_{\rm max}}= 2.5$ GeV
determine the density at $b \gtrsim 0.1$ fm with controlled uncertainties,
including the large-distance behavior where the function is exponentially small.
In this region the approach is completely data-driven and offers significant improvements
over model-dependent extractions.

The behavior of the density at $b \lesssim 0.1$ fm depends on qualitative assumptions
about the high-energy behavior of $\operatorname{Im} \FF (s)$. If the asymptotic sum
rule is disregarded, the logarithmic dispersion relation determines the density down
to much smaller $b \sim 10^{-3}$ fm, continuing the $\log (1/b)$ behavior observed
around $b \sim 10^{-1}$ fm. If the sum rule is enforced, the rise of the density is slowed;
however, the $\log \log (1/b)$ dependence in the pQCD asymptotic expression
is not visible at any practically relevant distances.

The status of the asymptotic sum rule for the pion FF remains unclear.
The empirical $\operatorname{Im} \FF (s)$ shows no sign of satisfying the asymptotic
sum rule at energies $s < s_{\rm max}$, even though the charge sum rule is well satisfied.
It is unlikely that experiments at slightly higher $s$ would change the situation.
Theory should revisit the principal assumptions in using pQCD approximations
in the dispersive approach, or pursue other explanations (e.g. possible heavy resonances).

The present analysis uses the $n = 1$ logarithmic dispersion relation
Eq.~(\ref{eq:DR2}).  It could eventually be extended using
the $n > 1$ relations, which reduce the sensitivity to high energies
but involve subtraction constants in the form of the derivatives of
the FF or its phase at threshold with high accuracy. 
The analysis could also be extended to the kaon electromagnetic FF, 
with a coupled-channel approach accounting for the $\pi\pi$ cut below 
the $K\bar K$ threshold.

{\it Acknowledgments}. CW acknowledges helpful suggestions by Peter Kroll and the late Michael Pennington,
and collaboration with Ina Lorenz on an earlier unpublished analysis
using logarithmic dispersion relations. ERA acknowledges discussions with Wojciech Broniowski.  

This material is based upon work supported by the U.S.~Department of Energy, 
Office of Science, Office of Nuclear Physics under contract DE-AC05-06OR23177.
ERA and PSP are partially funded by the Spanish Ministerio de Ciencia Innovaci{\'o}n 
y Universidades (MICIU/AEI /10.13039/501100011033 and ERDF/EU) under grants 
No. PID2020114767GB.I00 and PID2023.147072NB.I00. 
PSP is funded by Junta de Andaluc\'ia under the grant POSTDOC\_21\_00136 
and ERA and PSP under grant FQM225.  

\bibliography{pionphase}
\end{document}